\documentclass[aps,prl,twocolumn,showpacs]{revtex4}
\usepackage{graphicx}

\begin{document}


\title{Magnetron sputter deposition of a 48-member cuprate superconductor library: Bi$_2$Sr$_2$Y$_x$Ca$_{1-x}$Cu$_2$O$_{8+\delta}$ (0.5$\le x \le$1) linearly varying in steps of $\Delta x = 0.01$.}


\author{R.J. Sanderson and K.C. Hewitt}
\email{Kevin.Hewitt@dal.ca}
\homepage{http://www.physics.dal.ca/~hewitt/}
\affiliation{Dalhousie University, Department of Physics and
Atmospheric Science, Halifax, Nova Scotia, Canada B3H 3J5}
\date{\today}

\begin{abstract}
Using magnetron sputtering, a spatial composition spread approach
was applied successfully to obtain 48-member libraries of the
Bi$_2$Sr$_2$Y$_x$Ca$_{1-x}$Cu$_2$O$_{8+\delta}$ (0.5$\le$x$\le$1)
 cuprate superconducting system.
The libraries of each system were deposited onto (100) single
crystal MgO, mounted on a water cooled rotating table, using two
targets: the antiferromagnetic insulator
Bi$_2$Sr$_2$YCu$_2$O$_{8+\delta}$ (P=98 W RF) and the hole doped
superconductor Bi$_2$Sr$_2$CaCu$_2$O$_{8+\delta}$ (P=44 W DC). A low
chamber pressure of 0.81 mTorr argon is used to reduce scattering by
the process gas.  To minimize oxygen resputtering a substrate bias
of -20 V was used as well as a process gas free of oxygen. A rapid
thermal processor is used to post-anneal the amorphous deposited
films. A step annealing regime was used, with a ramp rate of 5
$^{\circ}$C/s for heating and cooling, with a first plateau at 780
$^{\circ}$C held for 200 s, and a second at 875 $^{\circ}$C held for
480 s. X-ray diffraction reveals that the films develop crystalline
order with the c-axis lattice parameter contracting linearly from
30.55 {\AA} (x=0.5) to 30.24 {\AA} (x=1.0) with increasing
Y-content, consistent with bulk values. The crystallized films are
polycrystalline, developing preferred orientation (c-axis parallel
to the substrate) for thinner members of the library. There is a
change of 0.01 in doping per library member which will enable
further studies to densely map phase space.

\end{abstract}


\maketitle

\section{Introduction}
It has been accepted for a long time that the temperature-hole
concentration phase diagram holds the key to our understanding of
the cuprate superconductors.  Researchers have primarily sampled
phase space discretely, using high quality single crystals, and have
identified a number of important features such as the
pseudogap\cite{timusk99} - a region of the phase diagram which is
marked by the supression of low energy excitations below a
temperature T*. Understanding the doping dependence of this feature,
especially near optimal doping (p=0.16) is thought to hold the key
to determining whether the pseudogap is a friend or foe of
superconductivity\cite{norman05}. That is, whether T* merges with
T$_c$, the superconducting onset temperature, or ends abruptly at
optimal doping is thought to determine, respectively, whether the
pseudogap phase is a precursor to, or competes with,
superconductivity. Answering this question requires a dense map of
phase space in order to trace the behaviour of T* and T$_c$.
Combinatorial materials science offers a method whereby such a dense
map of phase space can be obtained.

Combinatorial materials science methods produce various compositions
on a spatially addressable substrate.  In particular, one may
produce a continuous variation in composition across a substrate
using physical vapor deposition techniques.  This spatial
composition spread approach is therefore of benefit when one wishes
to map phase space densely.  Although CMS methods have been used to
show that particular superconducting phases can be
prepared\cite{xiang95}, to the author's knowledge the composition
spread approach has not been used to map phase space of the cuprates
and this report therefore represents the first step towards this
goal. In this paper, the spatial composition spread approach is
applied to the superconducting system
Bi$_{2}$Sr$_{2}$Y$_{x}$Ca$_{1-x}$Cu$_{2}$O$_{8+\delta}$, where $x$
varies quasi-continuously between 0.5 and 1.

\section{{Experimental}}
The composition spread approach uses simultaneous, magnetron
sputtering of two targets, e.g. A and B.  Masks are placed over
each target to produce a variation in the mass deposited onto a
rotating substrate as described in more detail in
reference\cite{dahn02}. When the mass varies linearly for one
target and is constant for the other, and the rotation speed is
sufficient to intimately mix the atoms, by an appropriate choice
of the powers applied to each target a film composition
A$_{1-x}$B$_x$ (0.5 $\le$ x $\le$ 1) is produced.   To produce
Bi$_2$Sr$_{2}$Y$_x$Ca$_{1-x}$Cu$_2$O$_{8+\delta}$ with
0.5$\le$x$\le$1, one may therefore co-sputter two targets:
A=Bi$_2$Sr$_2$CaCu$_2$O$_{8+\delta}$(Bi-Ca-2212)and
B=Bi$_2$Sr$_2$YCu$_2$O$_{8+\delta}$(Bi-Y-2212).

\subsection{Target preparation}
Targets of nominal composition (slightly enriched with Ca and Sr)
Bi$_{2.0}$Sr$_{2.05}$Ca$_{1.1}$Cu$_{2}$O$_{8+\delta}$ and
Bi$_{2.0}$Sr$_{2.05}$Y$_{1.1}$Cu$_{2}$O$_{8+\delta}$ are made
through the same three-stage solid-state reaction. Powders of
Bi$_2$O$_3$ (99.975\%, Alfa-Aesar), SrCO$_3$ (99.99\%, Alfa-Aesar),
CaO (99.95\%, Alfa-Aesar) and CuO (99.7\%, Alfa -Aesar) are mixed in
the appropriate stoichiometric ratios and ground together for two
hours using an agate auto grinder. The mixture is placed in an
Al$_{2}$O$_{3}$ crucible and reacted in air inside a Thermolyne
48000 box furnace.  The first reaction, which calcinate the powders,
is at 800$\,^{\circ}\mathrm{C}$ for 12 h (slow heat/slow cool at
$4\,^{\circ}\mathrm{C}$/min). After the first reaction, the powder
is ground for 2 h. The next two reactions are also in air and at a
temperature of $875\,^{\circ}\mathrm{C}$. In between these two
$875\,^{\circ}\mathrm{C}$ reactions the powder is ground for 2
hours. Once the target powders are synthesized, they are pressed
into pucks and hardened before use. To accomplish this, the powder
is ground manually with an agate mortar and pestle and sifted
through a 70 $\mu$m sieve.  Next, approximately 40 g of the sieved
powder is pressed into a 5.08 cm (2 in.) diameter by 0.5 cm thick
disc using a pressure of 13,000 psi. The disc is sintered for 30 hrs
at $875\,^{\circ}\mathrm{C}$ (Bi-Ca-2212) or 900 $^{\circ}$C
(Bi-Y-2212) to harden the target and reduce porosity.

\subsection{Film deposition}
 The film deposition apparatus is a Corona Vacuum Coaters V-37
sputtering system equipped with 5 magnetrons configured in a
side-sputtering arrangement (on-axis), where the substrate is
directly across (5.5 cm) from the target. To power the magnetrons
either an Advanced Energy MDX-1KDC supply or a combination of an
Advanced Energy RFX-600 generator and RTX-600 tuner is chosen for
the Bi-Ca-2212 (conductor) and Bi-Y-2212 (insulator) targets,
respectively.  The film was deposited onto a water-cooled, rotating
table (43 cm diameter) upon which two sets of three 1 inch x 1 inch
single crystal MgO (100) substrates (Superconductive Components
Inc.) were placed radially to cover the 75 mm sputtering track.   A
slotted aluminum mask, consisting of fifty-six 0.5 mm slots
separated by 1.52 mm, is placed over each set. The mask therefore
produces a 48-member library. Thus, the difference in composition
between each adjacent library member is $0.5/48 = 0.01$.  In front
of the Bi-Y-2212 target a mask was placed to produce a constant
profile of mass deposited on the substrate while the Bi-Ca-2212 had
a mask that produced a linear variation in mass deposited on the
substrate.  The chamber was pumped down to a base pressure of
3.9x10$^{-7}$ Torr, then an argon flow of 2 sccm was initiated to
create a process gas pressure of 0.81 mTorr. It is important to note
that \textbf{oxygen was not added to the process gas as it leads to
greater resputtering of the film}.  A bias of -20 V was applied to
each mask using a carbon brush assembly (detailed
elsewhere\cite{sanderson05}). The targets were co-sputtered while
the substrate table rotated at constant rate of 15 revolutions per
minute. The Bi-Ca-2212 target was powered by a DC supply operating
at 44 Watts, while the Bi-Y-2212 target was powered by an RF supply
operating at 98 W. These powers are chosen to produce equal
sputtering rates, in order to deposit the desired range of
compositions (0.5$\le$x$\le$1), where $x=[Y]/[Y+Ca]$.

\subsection{Post annealing}
Post-annealing using a Rapid thermal Processor RTP-600S (Modular
Process Technology Corp.) was performed on the amorphous
as-deposited films in dry air. A step annealing regime was found to
result in the best films. The regime starts with a ramp at 5 $^{o}$C
per second to 780$^{o}$C which is held for 200 seconds, followed by
another ramp a 5 $^{o}$C/sec to 875 $^{o}$C which is held for 480
seconds. The RTA is then cooled at 5 $^{o}$c/sec to room
temperature.

\subsection{Composition analysis}
Composition of the films was determined using energy dispersive
spectroscopy (EDS) measurements.  Al foil strips were taped onto
glass microscope slides and placed on the table. Following film
deposition, the Al foil strips were removed from the microscope
slides and affixed to an Al holder using double sided Cu tape. These
were analyzed for elemental composition using a  JEOL JXA-8200
Superprobe energy dispersive spectrometer equipped with a Noran
energy detector (0.133 keV energy resolution). A 7.0 kV electron
beam with a 50 nA current is used to analyze a 10 $\mu$m spot of the
film.
\subsection{Structural analysis}
For all deposited films, X-ray diffraction (XRD) spectra were
collected using an Inel CPS-120 with a curved position sensitive
detector. The  Cu$_{K_{\alpha1,\alpha2}}$ X-ray beam is incident
upon the sample at approximately 6 degrees and the curved position
sensitive detector collects all scattered X-rays from 2$\theta$ = 4
to 120 degrees. Collection time for the XRD spectra is 1800 seconds.
The Inel has computer controlled translation stages to scan samples
precisely and efficiently.

\subsection{Thickness measurement}
The film thickness is measured using a Veeco Dektak 8M, by dragging
a 12.5 $\mu m$ stylus with a force of 24.4 $\mu N$ over the film.
\section{Results}

\begin{figure}[htb]
\includegraphics[width=3.5in]{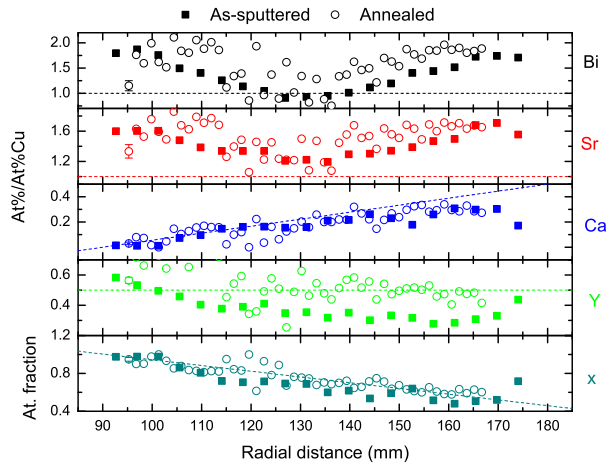}
\caption{Elemental composition (normalized to Cu) of
the library (SPO099), as-deposited (filled squares) and after
annealing (open circles).  Dashed lines indicate target
composition.} \label{eds}
\end{figure}

The composition of the films before and after post-annealing
treatment is shown in Fig. \ref{eds}

Figure \ref{eds} shows that post-annealing has no significant
effect on the elemental ratios, given an error of 5 \% in the
estimates.

\begin{figure}[htb]
\includegraphics[width=3.5in]{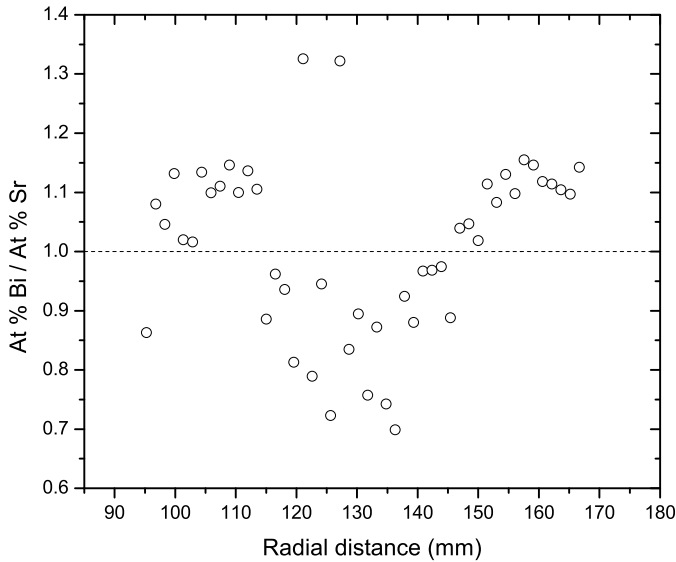}
\caption{The Bi:Sr elemental ratio of the
post-annealed library as deduced from the open circles in the two
upper panels of Fig. \ref{eds}.} \label{Bi-Srratio}
\end{figure}

The obvious question is whether the compositions in Fig. \ref{eds}
reflect the Bi2212 phase. While the Ca, Y content seems consistent
with expectations, the Bi:Cu and Sr:Cu ratios seem systematically
high, compared with the expected value (dashed line in Fig.
\ref{eds}). The first step is to determine the ratio of Bi:Sr, as
shown in Fig. \ref{Bi-Srratio}.

\begin{figure}[htb]
\includegraphics[width=3.5in]{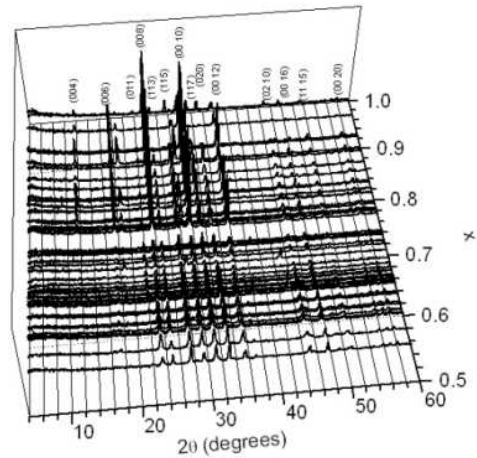}
\caption{X-ray diffraction spectra of library
(SP099), showing reflection consistent with pure Bi2212.  There is
one peak on the high angle side of the (006) reflection that we are
unable to identify.} \label{xrd}
\end{figure}

It is clear from this figure that most of the values are in the
range 0.7 to 1.15. It is known that Bi2212 can be formed for a
relatively large Bi:Sr range around 0.9-1.4 \cite{vanderah97}.
Therefore, one expects the structure to be crystalline Bi2212 except
perhaps in the range of radial distances 120-140 mm.  To determine
the structure, X-ray diffraction is carried out for the entire
library and the results are shown in the 3D plot of Figure
\ref{xrd}.

All reflections can be indexed to orthorhombic Bi2212 structure,
despite the range of Bi:Sr values seen in Fig. \ref{Bi-Srratio}. It
is surprising but fortuitous that the structure corresponds to
Bi2212 in the range 120-140 mm, given the Bi:Sr ratios.

\begin{figure}[htb]
\includegraphics[width=3.5in]{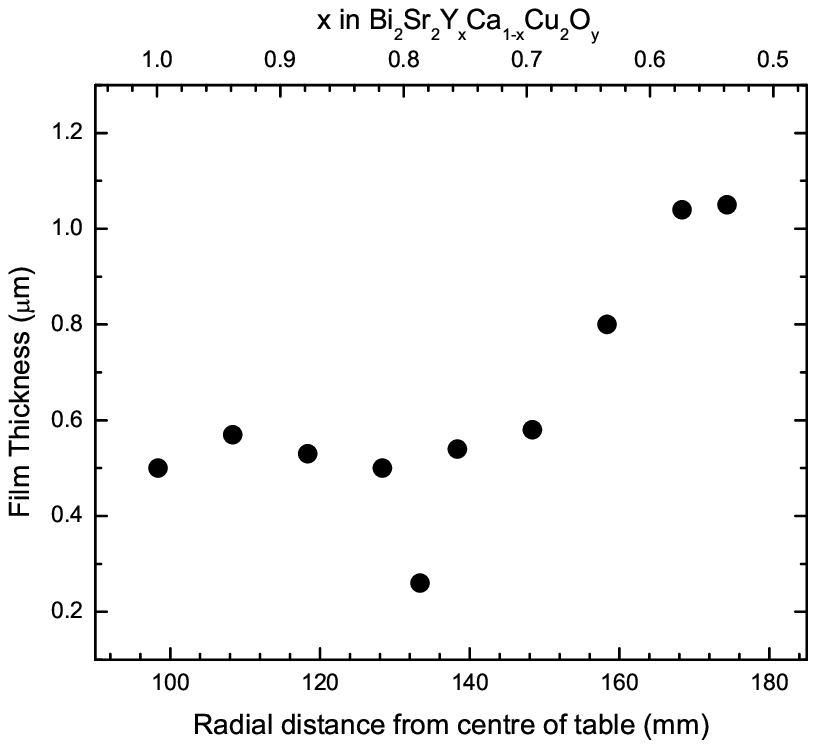}
\caption{The thickness at 10 points along the
library.} \label{thickness}
\end{figure}

Reflections from (00\emph{l}) planes dominate the XRD patterns,
leading to the conclusion that the film is primarily oriented with
the c-axis perpendicular to the substrate, especially for $x > 0.7$.
 To understand this behaviour the thickness of the library is
measured at 10 points and the results are shown in Figure
\ref{thickness}.

\begin{figure}[htb]
\includegraphics[width=3.5in]{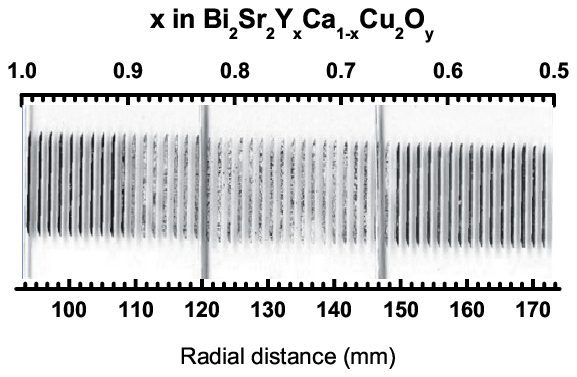}
\caption{Image of the 48-member library deposited on
(100) MgO.  There are actually 49 stipes but one of these falls
between two substrates near 148 mm and is not included in the XRD
analysis.} \label{libraryimage}
\end{figure}

The library is 0.5-0.6 $\mu m$ thick in the Y-rich range $x > 0.7$,
changing gradually to 1.0-1.1 $\mu m$ for Y-contents $x < 0.7$.  The
thickness does not change in a linear manner, which the linear plus
constant profile should produce.  However, as is evident in Fig.
\ref{eds}, there is some resputtering which would reduce the film
thickness in the central region of the library, as observed in Fig.
\ref{thickness}.  The library image of Figure \ref{libraryimage}
also shows
effects of resputtering, as the film also looks to be depleted in
the range 110-148 mm. It is known that epitaxial growth is
stabilized for very thin films. Thus, as expected, pseudo-epitaxial
growth is stabilized in the Y-rich region.

\begin{figure}[h!!tb]
\includegraphics[width=3.5in]{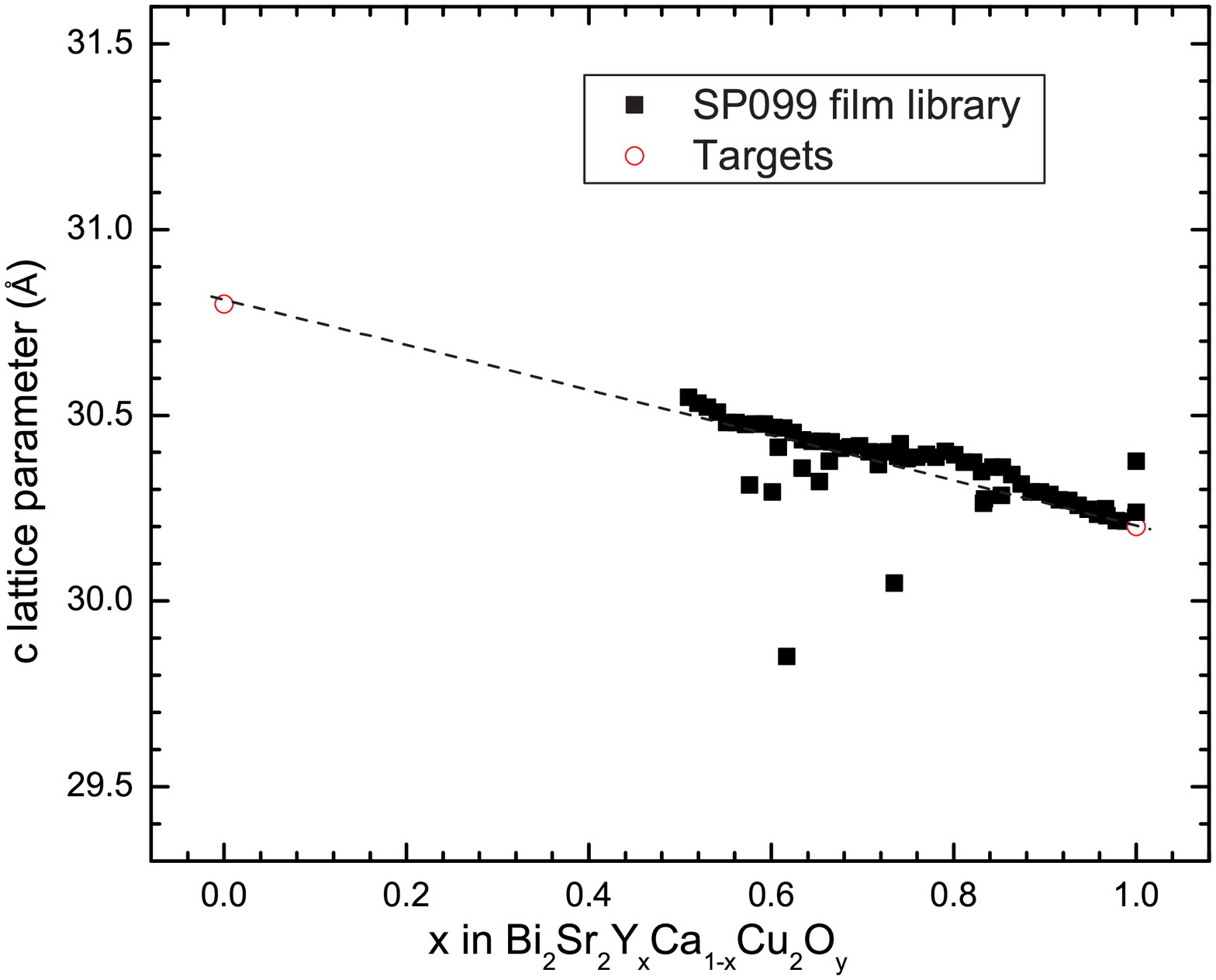}
\caption{The c-axis lattice parameter as a function of nominal
composition for the library (filled squares) and targets (open
circles).} \label{c-axis}
\end{figure}

Finally, the c-axis lattice parameter is determined in order to
ascertain whether the changes are consistent with those of bulk
Bi2212 doped with Y.  The results are plotted in Figure
\ref{c-axis}.
It reveals a trend that is clearly consistent with changes in bulk
Bi2212.

\section{Conclusions}
Structural and composition analysis has revealed that a spatial
composition spread approach has been successfully applied to the
cuprate superconductor system,
Bi$_2$Sr$_2$Y$_x$Ca$_{1-x}$Cu$_2$O$_{8+\delta}$ (0.5$\le$x$\le$1). A
48-member library is produced with the composition varying in a
linear manner over the range 0.5$\le$x$\le$1, corresponding to a
change in Y-content of 0.01 per library member.  The changes in the
lattice parameter are consistent with those of bulk, and preferred
orientation is observed for thinner members of the library.  These
films are now being used to densely map the electronic properties of
phase space.
\section{Acknowledgements}
The financial support of the Natural Sciences and Engineering
Research Council of Canada is gratefully acknowledged.  KH would
like to thank Ichiro Takeuchi for useful discussions.

\end{document}